\documentclass[referee]{aa} 

\usepackage{graphicx}
\usepackage{txfonts}
\usepackage{mathabx}
\usepackage[normalem]{ulem}

\def \RS{R$_{\odot}$}
\def \msol{M$\mathrm{_\odot}$}
\def \teff{$T_\mathrm{eff}$}

\def \1s{$1\,\sigma$}

\def \t0{T$_0$}

\def \aumic{AU~Mic}
\def \aumicb{AU~Mic~b}

\def \betapicb{$\beta$ Pic b}
\def \mpspG{m\,s$^{-1}$\,G$^{-1}$}
\def \mpspd{m\,s$^{-1}$\,d$^{-1}$}
\def \mps{m~s$^{-1}$}
\def \kps{km~s$^{-1}$}

\begin{document} 

   \title{Spin-orbit alignment and magnetic activity in the young planetary system AU Mic\thanks{ Based on observations obtained at the Canada-France-Hawaii Telescope (CFHT) which is operated from the summit of Maunakea by the National Research Council of Canada, the Institut National des Sciences de l'Univers of the Centre National de la Recherche Scientifique of France, and the University of Hawaii. Based on observations obtained with SPIRou, an international project led by Institut de Recherche en Astrophysique et Plan\'etologie, Toulouse, France.}}
   \author{ Martioli, E. 
            \inst{1,2} \and
    H\'ebrard, G.\inst{1,3} \and
    Moutou, C.\inst{4} \and
    Donati, J.-F.\inst{4} \and
    Artigau, \'E.\inst{5} \and
    Cale, B.\inst{6} \and
    Cook, N.J.\inst{5} \and
    Dalal, S.\inst{1} \and
    Delfosse, X. \inst{7} \and
    Forveille, T. \inst{7} \and
    Gaidos, E.\inst{9} \and
    Plavchan, P.\inst{6} \and
    Berberian, J. \inst{6} \and
    Carmona, A. \inst{7} \and
    Cloutier, R. \inst{13} \and
    Doyon, R.\inst{5} \and
    Fouqu\'e, P.\inst{8,4} \and
    Klein, B.\inst{4} \and
    Lecavelier des Etangs, A.\inst{1} \and
    Manset, N.\inst{8} \and
    Morin, J.\inst{10} \and
    Tanner, A.\inst{11} \and
    Teske, J. \inst{12} \and
    Wang, S. \inst{12}
    }
   \institute{
   \inst{1} Institut d'Astrophysique de Paris, UMR7095 CNRS, Universit\'e Pierre \& Marie Curie, 98~bis boulevard Arago, 75014 Paris, France, \email{martioli@iap.fr} \\
   \inst{2} Laborat\'{o}rio Nacional de Astrof\'{i}sica, Rua Estados Unidos 154, 37504-364, Itajub\'{a} - MG, Brazil\\
    \inst{3} Observatoire de Haute Provence, St Michel l'Observatoire, France \\
    \inst{4} Univ. de Toulouse, CNRS, IRAP, 14 avenue Belin, 31400 Toulouse, France \\
    \inst{5} Universit\'e de Montr\'eal, D\'epartement de Physique, IREX, Montr\'eal, QC, H3C 3J7, Canada\\
    \inst{6} George Mason University, 4400 University Drive, Fairfax, VA 22030, USA \\
    \inst{7} Univ. Grenoble Alpes, CNRS, IPAG, 38000 Grenoble, France\\
    \inst{8} Canada-France-Hawaii Telescope, CNRS, 96743 Kamuela, Hawaii, USA\\
    \inst{9} Department of Earth Sciences, University of Hawai’i at Mānoa, Honolulu, HI 96822, USA \\
    \inst{10} Universit\'e de Montpellier, CNRS, LUPM,34095 Montpellier, France\\
    \inst{11} Mississippi State University, Department of Physics \& Astronomy, Hilbun Hall, Starkville, MS, 39762, USA \\
    \inst{12} Observatories of the Carnegie Institution for Science, 813 Santa Barbara Street, Pasadena, CA 91101, USA \\
    \inst{13} Center for Astrophysics | Harvard \& Smithsonian, 60 Garden Street, Cambridge, MA, 02138, USA \\
    }
    
   \date{Received June XX, 2020; accepted July XX, 2020}

    \abstract{
    We present high resolution near-infrared spectropolarimetric observations using the SPIRou instrument at CFHT during a transit of the recently detected young planet \aumicb\,  with supporting spectroscopic data from iSHELL at IRTF. We detect Zeeman signatures in the Stokes V profiles, and measure a mean longitudinal magnetic field of $\overline{B}_\ell=46.3\pm0.7$~G. Rotationally modulated magnetic spots likely cause long-term variations of the field with a slope of $d{B_\ell}/dt=-108.7\pm7.7$~G/d. We apply the cross-correlation technique to measure line profiles and obtain radial velocities through CCF template matching. We find an empirical linear relationship between radial velocity and $B_\ell$, which allows us to estimate the radial velocity variations which stellar activity induces through rotational modulation of spots for the five hours of continuous monitoring of \aumic\ with SPIRou. We model the corrected radial velocities for the classical Rossiter-McLaughlin effect, using MCMC to sample the posterior distribution of the model parameters.  This analysis shows that the orbit of \aumicb\ is prograde and aligned with the stellar rotation axis with a sky-projected spin-orbit obliquity of $\lambda=0^{+18}_{-15}$ degrees. The aligned orbit of \aumicb\ indicates that it formed in the protoplanetary disk that evolved to the current debris disk around \aumic.
    } 
    

   \maketitle
%
\section{Introduction}
Detecting and characterizing planets around young stars is key to understanding the early stages of planetary evolution. Several mechanisms can produce strong misalignments between the planetary orbit and the stellar spin, including high eccentricity tidal migration, planet-planet scattering, and Kozai-Lidov cycles driven by a binary \citep[e.g.][]{DawsonAndJohnson2018, Triaud2018}. The resulting relative orientation of the planetary orbit and the rotation axis of the host star is a key discriminant between different formation and migration scenarios.  

Here we report a measurement of the spin-orbit angle for the recently detected transiting super-Neptune planet \aumicb\ \citep{plavchan2020Natur}. \aumic\ is a young and active M1 star with a spatially resolved edge-on debris disk \citep{kalas2004}, and a member of the $\beta$~Pictoris Moving Group \citep{torres2006}. Its distance of only $9.7248\pm0.0046$~parsec \citep{gaiadr22018} and its estimated age of $22\pm3$~Myr \citep{mamjek2014} make it both the closest and the youngest system with either a spatially resolved edge-on debris disk or a transiting planet. Table \ref{tab:aumicstarparams} summarizes the stellar and planetary parameters of the system. 

Young systems with detected planets \citep[e.g., V830 Tau b;][]{donati2016}, and especially those with either a remnant debris disk like \betapicb\ \citep{Lagrange2009} or transiting planets \citep[e.g., K2-33 b, DS Tuc Ab;][]{David2016, Mann2016, Newton2019} are key probes of planetary formation.  \aumic\ has both a disk and at least one transiting planet, and it is also unique among debris disk hosts for being an M star, the most numerous type of star in our Galaxy and the most promising spectral type to find habitable planets using current techniques.  

\begin{table}
\centering
\caption{Star and planet b parameters for \aumic\ system}
\label{tab:aumicstarparams}
\begin{tabular}{ccc}
\hline
Parameter & Value & Ref. \\
\hline
\teff & $3700\pm100$~K & 1 \\
Star mass  & $0.50\pm0.03$~\msol & 1 \\
Star radius  & $0.75\pm0.03$~\RS & 2 \\
P$_{\rm rot}$  & $4.863\pm0.010$~days & 1 \\
$v_{e}=2\pi R_{\star} / {\rm P}_{\rm rot}$ & $7.8\pm0.3$~\kps  & 2,1 \\
Age  & $22\pm3$~Myr & 3 \\
Distance  & $9.7248\pm0.0046$~parsec & 4 \\
Limb dark. coef. $\mu_{H}$  & 0.3016 & 5 \\
\hline
Time of conjunction & $2458330.39153^{+0.00070}_{-0.00068}$~BJD & 1\\
Transit duration & $3.50^{+0.63}_{-0.59}$~hr & 1\\
Orbital period      & $8.46321\pm0.00004$~days   & 1 \\
RV semi-amplitude & $<28$~\mps & 1\\
$R_{p}/R_{\star}$ &  $0.0514\pm0.0013$ & 1 \\
$a_{p}/R_{\star}$  &  $19.1^{+1.8}_{-1.6}$  & 1\\
Impact parameter ($b$)  & $0.16^{+0.14}_{-0.11}$ & 1 \\
Orbit inclination ($i_{p}$)  & $89.5^{\circ}\pm0.4^{\circ}$ & 6\\
\hline
\end{tabular}
\tablebib{
(1) \citet{plavchan2020Natur}; 
(2) \citet{russel2015};
(3) \citet{mamjek2014}; 
(4) \citet{gaiadr22018}; 
(5) \citet{claret2011};
(6) $\cos{i_{p}} = \frac{b}{a_{p}/R_{\star}}$
}
\end{table}

\section{Observations and data reduction}
\label{sec:dataacqred}

\subsection{SPIRou}

The Spectro-Polarim\`etre Infra Rouge (SPIRou)\footnote{more information about SPIRou in \url{http://spirou.irap.omp.eu} and \url{https://www.cfht.hawaii.edu/Instruments/SPIRou/}} is a stabilized high resolution near infrared spectro-polarimeter \citep[submitted]{donati2020} mounted on the 3.6~m Canada-France-Hawaii Telescope (CFHT) atop of Maunakea, Hawaii. SPIRou is designed to perform high precision measurements of stellar radial velocities to search and characterize exoplanets.  It provides full coverage of the near infrared spectrum from 950~nm to 2500~nm, in a single exposure, without gaps, and at a spectral resolving power of $\lambda / \Delta\lambda \sim 70000$. Its high throughput in the near infrared makes SPIRou an ideal instrument to follow-up transiting exoplanets around cool stars. SPIRou allows simultaneous spectropolarimetry, which helps identify stellar magnetic activity and is especially important for active late type stars \citep{Morin2010} and young stellar objects. \aumic\ is both cool and young, with a high magnetic activity \citep{Berdyugina2008, Afram2019} which requires polarimetric information to reliably diagnose.

\subsection{Observations}

We observed the June 16, 2019 transit of \aumicb\ as part of the Work Package 2 (WP2) of the SPIRou Legacy Survey \citep[submitted]{donati2020} CFHT large program (id 19AP42, PI: Jean-Fran\c{c}ois Donati). The observations were carried out in the Stokes~V spectropolarimetric mode of SPIRou. They started at UT 2019-06-17T10:10:56 and finished at UT 2019-06-17T15:13:45, and consist of 116 individual flux spectra of \aumic\ with a 122.6~s exposure time. They corresponds to 29 Stokes~V polarimetric spectra (with $4\times$ individual exposures per polarimetry sequence). Our observations started with an air mass of 2.9 and ended at 1.8, with a minimum of 1.59. The conditions remained nearly photometric, with the SkyProbe monitor \citep{Cuillandre2004} measuring a maximum absorption of 0.12 mag.  The image quality (seeing) measured by the SPIRou guider varies from 0.8 to 1.6~arc seconds, with a mean value of $0.96\pm0.13$~arc second. The Moon was almost full, with a 99\% illumination, and was separated from our target by 40.3 degrees. The peak signal-to-noise ratio (SNR) per spectral bin (in the spectral order centered at $\sim1670$\,nm) of the individual exposures varies between 176 to 273, with a mean value of 242. 

\subsection{Data reduction}

The data have been reduced with version v.0.6.082 of the  APERO SPIRou data reduction software (Cook et al, 2020, in prep.).  APERO first combines the sub-exposures at the read-out level, correcting for the non-linearity in the pixel-by-pixel response. The 1D spectral fluxes are optimally extracted following \cite{horne86}. The individual spectral orders are processed and saved separately, providing a 2D frame with about 4088 spectral pixels for 48 orders. SPIRou uses two optical fibers to collect light from the two images formed by a Wollaston prism. For pure spectroscopy, APERO merges the spectra of the two beams, whereas for polarimetry the fluxes of the two channels are individually saved for later polarimetric analysis.  APERO also calculates for each channel a blaze function from daytime flat-field exposures. The pixel-to-wavelength calibration is obtained from a combination of daytime exposures of a UNe hollow-cathod lamp and of a thermally-controlled Fabry-P\'erot etalon (FP). The FP also feeds a third fiber during science exposures to monitor instrument drifts. APERO calculates a telluric absorption spectrum for each exposure, using an extensive library of telluric standard stars observed nightly with SPIRou over a wide range of air mass and atmospheric conditions. APERO uses the PCA-based correction technique of \cite{artigau14} to produce a telluric-absorption corrected spectrum. APERO also calculates the Cross Correlation Function (CCF) with a set of line masks optimized for different stellar types and systemic velocities.

\subsection{iSHELL data}

We include in our analysis simultaneous RV measurements from 47 in-transit spectra of the June 16, 2019 transit of \aumic\ obtained with the iSHELL spectrometer ($\lambda / \Delta\lambda \sim$80,000) on the NASA Infrared Telescope Facility \citep[IRTF,][]{Rayner2016}. AU Mic was observed in KGAS mode (2.1 - 2.5~$\mu$m) from UT 2019-06-17T11:08:19 to UT 2019-06-17T12:53:32. Their 2-minute exposure time results in a photon signal-to-noise ratio of $\sim$ 60-70 per spectral pixel at $2.4\,\mu$m (the approximate peak of the blaze function for the center order), and in turn in a RV precision of 15-27~\mps\ (median 21~\mps) per measurement. These spectra were reduced and their RVs extracted using the methods outlined in \citet{Cale2019}. The RV data measured by iSHELL are presented in Appendix \ref{app:rvdata}.

\section{Spectropolarimetry}
\label{sec:spectropolarimetry}

SPIRou Stokes-V spectra are obtained from sequences of 4 exposures with distinct positions of the Fresnel rhombs such that systematic errors affecting the polarimetric analysis are minimized (we compute the Stokes parameter using the "ratio" method \citep{donati97, bagnulo2009}). Since the order of the exposures within the successive \aumic\ polarimetric sequences is identical and the angles of the retarder within each sequence are set to alternate positions, one can obtain higher time sampling by calculating polarimetric spectra in every set of four adjacent exposures. With this method we obtain a total of 113 (non-independent) polarimetric spectra of \aumic\, instead of the 29 that would be obtained from analyzing each sequence separately. 

We applied the least squares deconvolution (LSD) method of \cite{donati97} to each Stokes I, Stokes V, and null polarization spectrum to obtain LSD profiles for each. The line mask used in our LSD analysis was obtained from the VALD catalog \citep{piskunov1995} based on a MARCS model atmosphere \citep{Gustafsson2008} of effective temperature 3500~K and surface gravity $\log{g}=5.0$~cm~s$^{-2}$. A total of 1363 lines was included in the LSD analysis.

\begin{figure}
   \centering
   \includegraphics[width=0.9\hsize]{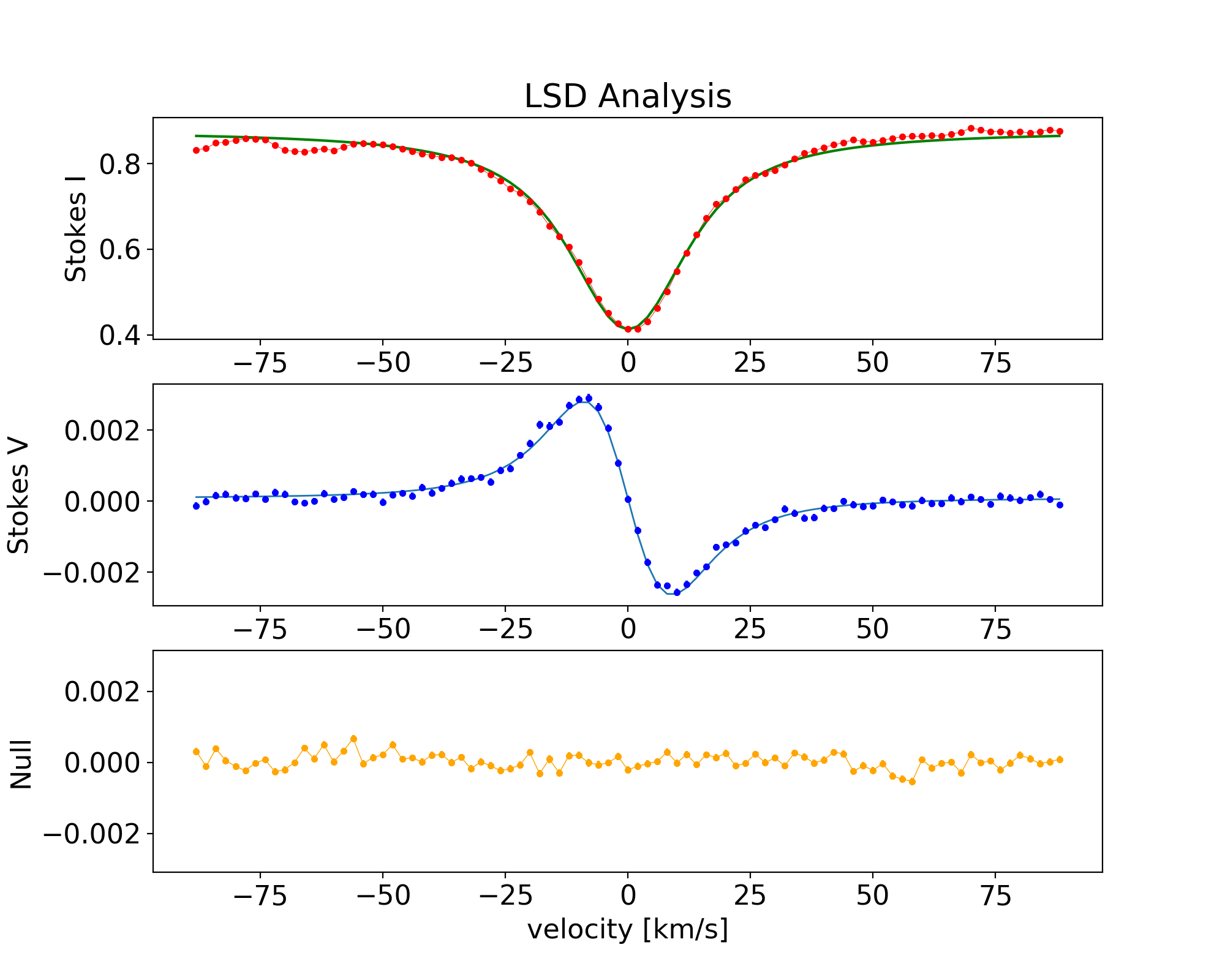}
      \caption{Median of all LSD profiles in the \aumic\ time series. The top panel shows Stokes I LSD (red points) with a 
       a Voigt profile model fit (green line); the middle panel shows Stokes V (blue points) and a double Voigt profile 
       model fit (blue line);  the bottom panel shows the null polarization profile (orange points). 
       }
        \label{fig:spirou-lsd-profiles}
  \end{figure}

Fig. \ref{fig:spirou-lsd-profiles} presents the medians of the 113 profiles, and its Stokes-V panel shows a clearly detected Zeeman signature. We fit a Voigt function to the Stokes-I profile and a double Voigt function to the Stokes-V profile, both presented in Fig. \ref{fig:spirou-lsd-profiles}. The Voigt model is a good approximation for the profiles of \aumic, which confirms a significant contribution from Lorentzian broadening mechanisms, most likely due to its high surface gravity. A complete analysis of the line profiles considering the several broadening mechanisms in \aumic\ is out of the scope of this paper.  The fit profiles are also important in this work to correct for the velocity shift in the profiles, which is needed for the calculation of the longitudinal magnetic field as given by Eq. \ref{eq:Bl} in Sec. \ref{sec:longmagfield}.

\section{Radial Velocities}
\label{sec:radialvelocities}

We measure the radial velocity of \aumic\ using the cross-correlation function (CCF) between the telluric-corrected stellar spectrum and a line mask \citep{pepe2002}.  The broad nIR band pass of SPIRou covers thousands of atomic and molecular lines, which greatly improves the precision in the determination of the CCF. The line mask plays an important role in the CCF method, since it determines the spectral regions that are probed and the statistical weight for each of these regions. We use the ``M2\_weighted\_RV\_-5.mas'' line mask from the set of empirical masks delivered by the APERO pipeline. This mask is based on the observed spectra of the M2V star Gl~15A and is a good match to the M1V spectral type of \aumic. 

Even though SPIRou spectra are corrected for telluric absorption, this correction can create additional noise in the spectrum. This noise has been evaluated using SPIRou commissioning data, and it has been taken into account in the weight of each line in the mask, which is decreased by a factor proportional to the extra-noise. The lines are blanked out if they are impacted by telluric deeper than 40\% absorption at a systemic radial velocity of -5~\kps\ (which is close to the 4.5~\kps\ systemic velocity of \aumic) with a window of $\pm33$~\kps\ (maximum of barycentric velocity). For a telluric absorption of 10\%, 20\%, 30\% and 40\%, the weights are given by the line depth divided by a factor of 1.5, 3, 7, and 16, respectively.

The mask has 3475 lines, but we further filter it using the approach of \citet[in press]{moutou2020}, eliminating those lines which are not detected in the mean Stokes-I spectrum of \aumic, for a final set of 2277 retained lines. Notice that one could have obtained the radial velocities from the LSD Stokes-I profiles as presented in Sec. \ref{sec:spectropolarimetry}.  However our LSD analysis is restricted to spectral lines with known Land\'{e} factor, which is smaller compared to the number of lines in the CCF analysis, resulting in larger uncertainties in radial velocities.

The 48 orders delivered by SPIRou have different noise levels, depending mostly on the instrumental throughput \citep[submitted]{donati2020} and on the telluric absorption. We compute a separate CCF for each spectral order, and combine some of those into a sum CCF to improve precision. We obtain individual RV measurements for each spectral order and calculate the RV dispersion $\sigma_{\rm RV}$, given by the standard deviation throughout the time series. The mean RV dispersion between all orders is $\overline{\sigma}_{\rm RV}=97\pm90$~\mps.  Given the variable RV precision between orders, we decided to restrict our analysis to the seven orders in the 1512~nm to 1772~nm range in the H-band, where the mean RV dispersion is $\overline{\sigma}_{RV}=28\pm7$~\mps.  Our CCF mask has a total of 842 lines within this spectral range. 

We measure radial velocities from the CCF by least-square fitting for the velocity shift $\Delta v_{i}$ that best matches the CCF of an individual exposure, CCF$_{i}$, to the median of the CCFs of all exposures, CCF$_{m}$. The shifted template CCF$_{m}(v +\Delta v)$ is calculated by cubic interpolation.  We also measured RVs by fitting a Gaussian to each CCF$_{i}$, which is the most usual method.  That gives similar results but shows stronger systematics correlated with the air mass of the observations, and we therefore adopt the CCF matching (CM) method in our analysis. In yet another processing alternative, we apply a median filter (MF) to the CCF time series before calculating RVs through template matching, using a 3 x 3 window along the time and velocity axes. The RV data measured by SPIRou are presented in Appendix \ref{app:rvdata}.


\section{Rossiter-McLaughlin effect}
\label{sec:classicalrmeffect}

We first model the SPIRou radial velocities of \aumic\ obtained from the median-filtered CCFs as the combination of its reflex orbital motion caused by planet b, assuming a circular orbit and the \cite{plavchan2020Natur} orbital parameters, and the classical Rossiter-McLaughlin (RM) effect, with the stellar limb darkening accounted for as described in \citet{ohta2005}. We adopt a linear limb darkening model and fix the H-band coefficient to $\mu_{H}=0.3016$ from \cite{claret2011}. 

We adopt as free parameters the time of conjunction $T_{c}$, the planet to star radius ratio $R_{p}/R_{\star}$, the sky projected obliquity angle $\lambda$, the projected stellar rotation velocity $v_{e}\sin i_{\star}$, the systemic velocity $\gamma$, and we include a slope of the RVs as a function of time, $\alpha$, to account for both stellar activity trends and a planetary signal.  Table \ref{tab:aumicfitparams} shows the priors which we adopt for each parameter. We sampled the posterior distributions using the \texttt{emcee} Markov chain Monte Carlo (MCMC) package \citep{foreman2013}, using 50 walkers and 2000 MCMC steps of which we discard the first 500. The best-fit values in Table \ref{tab:aumicfitparams} are the medians of the posterior distribution, and the error bars enclose 34\% on each side of the median. The MCMC samples and posterior distributions are illustrated in Fig. \ref{fig:rmfit-to-spirourv_pairsplot} in Appendix \ref{app:posteriordistributions}.

\begin{table}
\centering
\caption{Fit parameters of \aumicb. T$_{c}$ is in units of BJD - 2458330, $\lambda$ is in degrees, $v_{e}\sin i_{\star}$ and $\gamma$ are in \kps, and $\alpha$ is in \mpspd. The symbol $\mathcal{N}(\mu,\sigma)$ represents a normal distribution with mean $\mu$ and standard deviation $\sigma$, and $\mathcal{U}(x_{1},x_{2})$ represents a uniform distribution with minimum and maximum values given by $x_{1}$ and $x_{2}$.}
\label{tab:aumicfitparams}
\begin{tabular}{ccc}
\hline
Parameter & Prior & Posterior \\
\hline
T$_{c}$ & $\mathcal{N}(0.39153,0.00070)$ & $0.3892\pm{0.0007}$ \\
$R_{p}/R_{\star}$ & $\mathcal{N}(0.0514,0.0013)$ & $0.063\pm{0.004}$ \\
$\lambda$ & $\mathcal{U}(-180,180)$ & $-0.2^{+18.9}_{-19.3}$ \\
$v_{e}\sin i_{\star}$ & $\mathcal{N}(7.8,0.3)$ &  $7.5\pm{0.9}$\\
$\gamma$ &  $\mathcal{U}(-\infty,\infty)$ & $-4.3869\pm{0.0005}$\\
$\alpha$  & $\mathcal{U}(-\infty,\infty)$ & $149\pm{9}$\\
\hline
\end{tabular}
\end{table}

\begin{figure*}
   \centering
   \includegraphics[width=0.9\hsize]{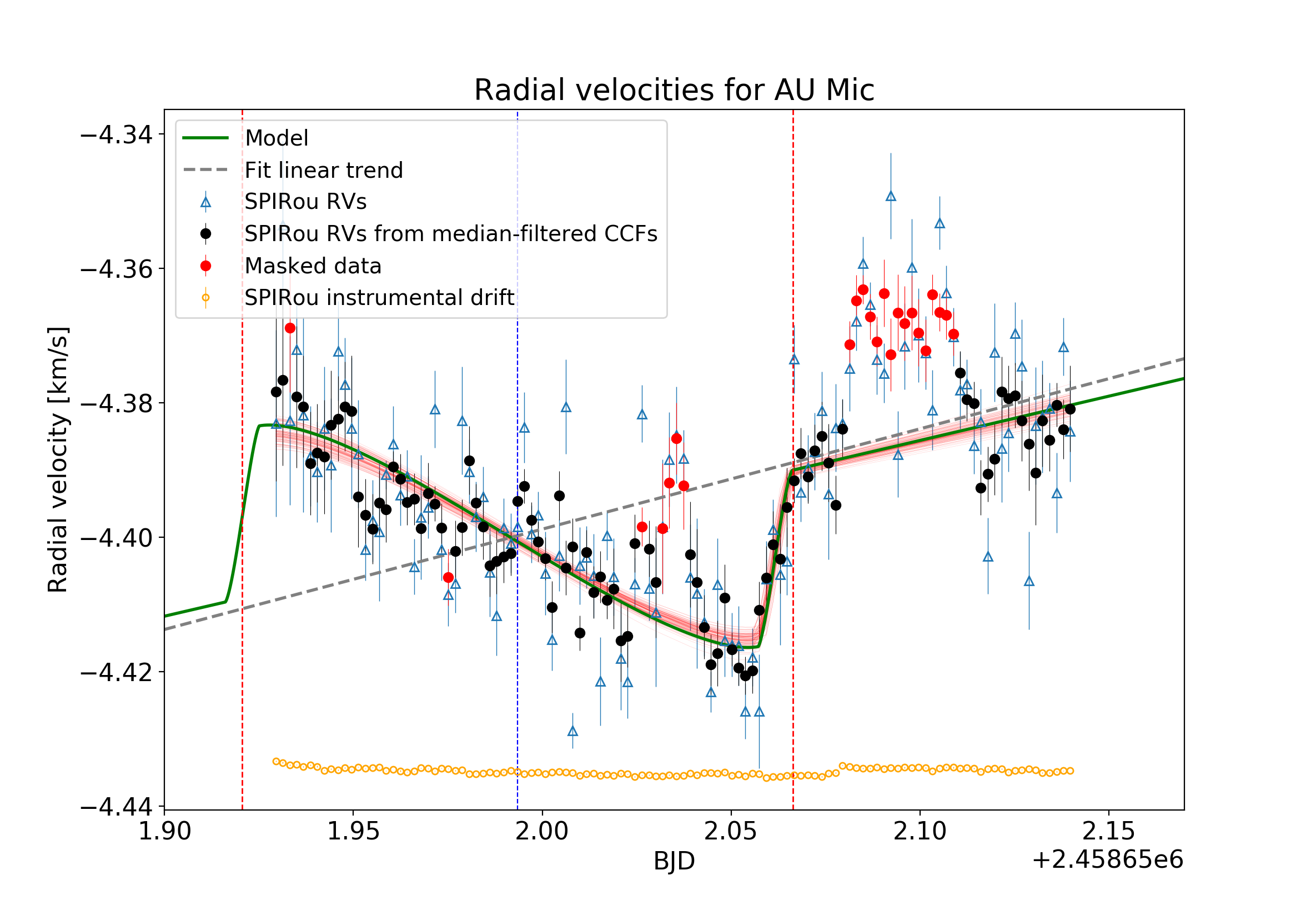}
      \caption{ SPIRou radial velocities of \aumic. Blue triangles show the RVs obtained 
        from CCF matching the original CCFs, and filled circles show the RVs obtained  
        from CCF matching the median-filtered CCFs. Red circles show data masked by our 
        2.5$\sigma$ clip. Vertical lines show the predicted start, center, and end of 
        the transit. The green line shows the best fit model and the thin red lines 
        show models for 100 randomly selected MCMC samples. The gray dashed line shows 
        best fit slope of $149\pm{9}$~\mpspd\ with an arbitrary vertical offset for visualization, 
        and the orange points show the SPIRou instrumental drift (also with an arbitrary offset), 
        and illustrate its dispersion of just $0.51$~\mps.
        }
        \label{fig:rmfit-to-spirourv}
  \end{figure*}

Fig. \ref{fig:rmfit-to-spirourv} shows as blue triangles the SPIRou \aumic\ RVs obtained by CCF matching the original CCFs, whereas the filled circles show those obtained from the median filtered CCFs. We identify two anomalous regions in the time series, marked in red in the figure, where the RV residuals are above $2.5\times\sigma$. We interpret these regions as stellar activity events, such as spot-crossing by the planet and/or flares. The corresponding data were masked out in the final model fit. Our best fit RM model includes a RV slope of $149\pm{9}$~\mpspd\ and the dispersion of its residuals is 5.1~\mps\ for data that were not masked out. For illustration of the stability of SPIRou, we also show the instrumental drifts obtained from the spectrum of the FP calibrator which is simultaneously observed through the reference fiber, with a dispersion of just $0.51$~\mps.

The sky projected obliquity angle of $\lambda=-0.2^{+18.9}_{-19.3}$ degrees shows that the orbit of \aumicb\ is prograde and close to aligned with the rotation axis of the parent star. Our best fit value of $v_{e}\sin i_{\star}=7.5\pm{0.9}$~\kps\ agrees at level of $2\times\sigma$ with independent measurements of $v_{e}\sin i_{\star}=8.7\pm0.2$~\kps \citep{gaidos2014}. Our analysis also shows that the conjunction occurred about 3.4 minutes ($\sim3\times\sigma_{T_c}$) earlier than predicted, and favors a slightly larger planetary radius, though within $3\sigma_{R_{p}/R_{\star}}$.

\section{Magnetic activity}
\label{sec:longmagfield}

As amply illustrated by its TESS light curve, \aumic\ is an active star, with a surface largely filled by spots, and with frequent flares \citep{plavchan2020Natur}. The $\sim5$~hr SPIRou time series covers 4.3\% of the 4.863-day rotation period of \aumic\ . The non-uniform brightness distribution of the \aumic\ disk has therefore probably changed, slowly through rotation of the visible hemisphere, and rapidly through flaring and spot evolution. Planet \aumicb\ can additionally transit spots, also causing fast variability.  These brightness variations change the rotation profile of \aumic\ and strongly affect our RV measurements.  

Since both spot and flare events are connected to the magnetic field \citep{Lavail2018}, we search for an empirical correlation between the measured RVs and the longitudinal magnetic field $B_\ell$, in an attempt to mitigate the effects of stellar activity on our RV data. The longitudinal magnetic field $B_\ell$ is calculated for each \aumic\ polarized spectrum using the following equation from \cite{donati97}:

\begin{equation}
    \label{eq:Bl}
    B_\ell=-2.14 \times 10^{11} \frac{\int vV(v)\mathrm{d}v}{\lambda_0 \cdot g_{\rm eff} \cdot c \cdot \int \left[ I_c - I(v) \right] \mathrm{d}v}
\end{equation}

where $c$ is the speed of light, $I(v)$ and $V(v)$ are the Stokes I and V profiles as functions of velocity $v$ in the star's frame, $I_c$ is the continuum of the Stokes I profile, $\lambda_0=1515.38$~nm is the mean wavelength, and $g_{\rm eff}=1.24$ is the mean Land\'e factor of the lines included in the mask. The $B_\ell$ data are provided in Appendix \ref{app:Bldata}. The bottom panel of Fig. \ref{fig:bl_rv_fit} illustrates our measurements of $B_\ell$ for \aumic, showing values obtained both from the original Stokes V profiles (black points with error bars) and from the median filtered profiles (black line). The mean longitudinal field of \aumic\ during our time series is $\overline{B}_\ell=47.9\pm8.1$~G, with a linear trend of slope $-108.7\pm7.7$~G/d which is likely due to rotational modulation of spots. The field measured from the median LSD profile of Fig. \ref{fig:spirou-lsd-profiles} is $\overline{B}_\ell=46.3\pm0.7$~G and therefore closely matches the mean longitudinal field of the sequence.  

We least-square fit (Fig. \ref{fig:bl_rv_fit}, top panel) the following linear function to the RM-subtracted RV data:

\begin{equation}
    \label{eq:rvmodelfromBl}
    v_{B}(t) = [B_\ell(t) - B_0] \, a + v_{0}, 
\end{equation}

where $B_{0}$ is an arbitrary reference magnetic field, $a$ is the scaling factor between the two quantities, and $v_{0}$ is a constant velocity. The best fit scaling factor is $a=-1.34\pm0.12$~\mpspG, significant at the 11$\sigma$ level. The Pearson-$r$ correlation coefficient between our measured RVs and the predicted $v_{B}$ is $r=0.72$ with a p-value of $3.7\times10^{-19}$, showing a significant correlation between the two quantities, mainly because stellar rotation modulates both the RVs and $B_\ell$. Subtracting a linear fit from both $B_\ell$ and RVs to eliminate the long term variations reduces $r$ to $0.19$ with a p-value of $4.6\times10^{-2}$, showing some possible smaller correlation between the short time-scale variations of the RVs and $B_\ell$ . Subtracting only the fitted $B_\ell$ slope however produces less dispersed RV residuals than subtracting the full empirical model $v_{B}$.  The short time scale structure is likely due to spot evolution, flares, and the planet transiting spots.  Each of these phenomena unfortunately has a different relationship between its RV variation and $B_\ell$, which makes our linear model much too simple to account for the short-term RV variability. A future paper will investigate these issues in much more detail and with an extended observational dataset.

  \begin{figure}
   \centering
   \includegraphics[width=0.9\hsize]{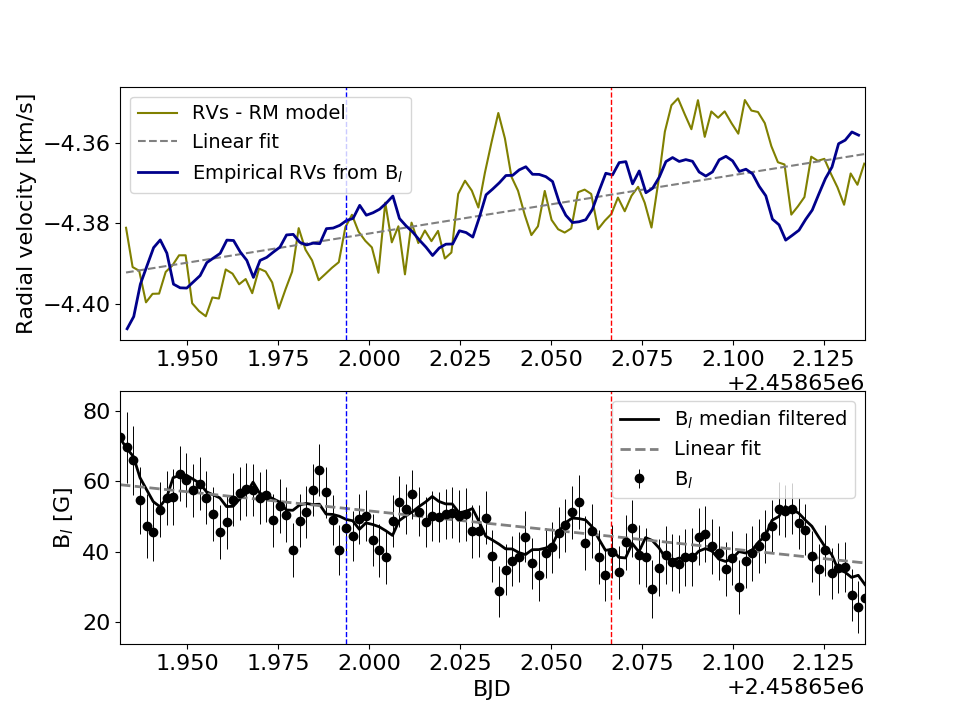}
      \caption{The top panel shows as a function of time the RM-subtracted RVs of \aumic\ in green and 
        their best fit Eq. \ref{eq:rvmodelfromBl} linear model in dark blue. The vertical dashed 
        lines show the predicted transit center (blue) and end (red). The bottom panel shows the longitudinal 
        field derived from the original LSD profiles (black circles) and from the median filtered LSD profiles 
        (black line). We also present a linear fit to the values of $B_\ell$ (dashed grey line in bottom panel) 
        and the corresponding trend in velocity space (dashed gray line in top panel). 
              }
        \label{fig:bl_rv_fit}
  \end{figure}

\section{Results and Discussion}
\label{sec:results}

Our preferred SPIRou RVs of \aumic\ are obtained by subtracting from the measured RVs the linear component of the empirical model, a $145\pm{17}$~\mpspd\ slope which mostly removes the stellar activity signal discussed above. We then adjust the RM model of Sec. \ref{sec:classicalrmeffect} to both the iSHELL and corrected SPIRou data, using 50 MCMC walkers and 2000 steps with the first 500 discarded. We consider two systemic velocities $\gamma_{\rm SPIRou}$ and $\gamma_{\rm iSHELL}$, to account for different instrumental zero points. The MCMC samples and posterior distributions are illustrated in Fig. \ref{fig:rmfit-to-spirou+ishell_pairsplot} in Appendix \ref{app:posteriordistributions}.

\begin{figure}
   \centering
   \includegraphics[width=0.9\hsize]{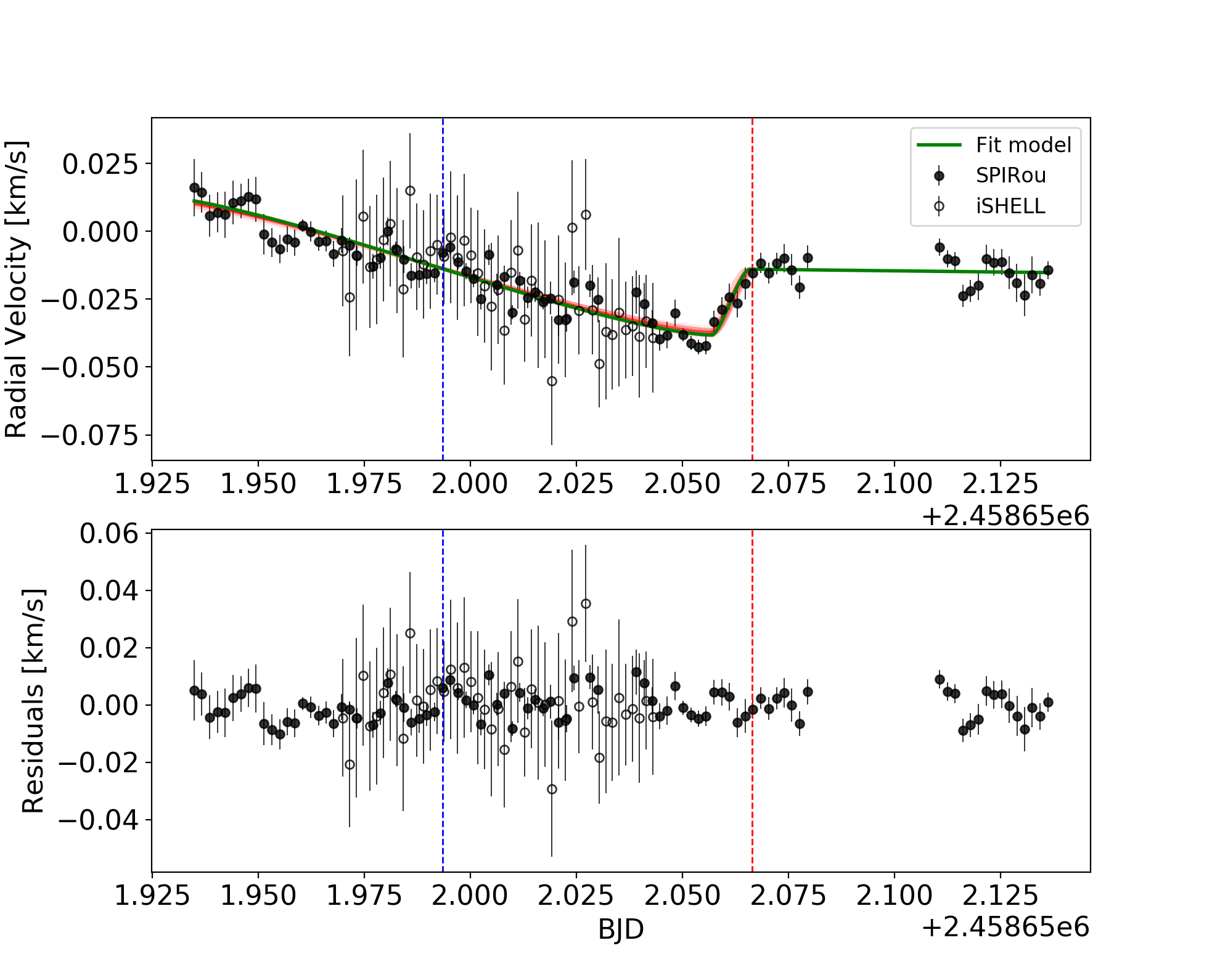}
      \caption{Simultaneous fit to the corrected SPIRou RVs (filled circles) and iSHELL RVs 
       (hollow circles) of the model of Rossiter-McLaughlin effect (green line shows the
       best fit model and the thin red lines show models for 100 randomly selected MCMC samples).
       The vertical dashed lines show the predicted transit center (blue) and end (red). Bottom 
       panel shows the residuals of the fit with respective dispersions of 5.1~\mps\ and 
       11.5~\mps\ for SPIRou and iSHELL. }
        \label{fig:rmfit-to-spirou+ishell}
 \end{figure}

The final fit parameters are presented in Table \ref{tab:aumicfinalfitparams}. We obtain a fitted obliquity angle of $\lambda=-0.3^{+17.8}_{-15.0}$ degrees and 5.1-\mps\ and 11.5-\mps\ dispersions for the SPIRou (masked data excluded) and iSHELL residuals. This result confirms that the planet is on a prograde orbit and that the orbital and rotation spins are closely aligned. Fig. \ref{fig:rmfit-to-spirou+ishell} shows this final fit model to the RV data for both instruments.

\begin{table}
\centering
\caption{Final fit parameters for the Rossiter-McLaughlin model of \aumicb\ using both SPIRou and iSHELL data sets. T$_{c}$ is in units of BJD - 2458330, $\lambda$ is in degrees, $v_{e}\sin i_{\star}$, $\gamma_{\rm SPIRou}$ and $\gamma_{\rm iSHELL}$ are in \kps.}
\label{tab:aumicfinalfitparams}
\begin{tabular}{ccc}
\hline
Parameter & Prior & Posterior \\
\hline
T$_{c}$ & $\mathcal{N}(0.39153,0.00070)$ & $0.3897\pm{0.0006}$ \\
$R_{p}/R_{\star}$ & $\mathcal{N}(0.0514,0.0013)$ & $0.061\pm{0.002}$ \\
$\lambda$ & $\mathcal{U}(-180,180)$ & $-0.3^{+17.8}_{-15.0}$ \\
$v_{e}\sin i_{\star}$ & $\mathcal{N}(7.8,0.3)$ & $7.8\pm{0.6}$\\
$\gamma_{\rm SPIRou}$ &  $\mathcal{U}(-\infty,\infty)$ &  $-4.3808\pm{0.0005}$ \\
$\gamma_{\rm iSHELL}$  & $\mathcal{U}(-\infty,\infty)$ &  $0.027\pm{0.003}$ \\
\hline
\end{tabular}
\end{table}

Since iSHELL only observed a partial transit of \aumic\ and no out-of-transit baseline, its data alone do not constrain a full RM model independently of SPIRou, but the two data sets are fully mutually compatible. The agreement between the data sets from these two different instruments using independent techniques for data analysis is remarkable and shows that both instruments are stable and can provide RVs with precisions of a few \mps\ for an active star.

In addition to the analysis presented here, we performed extensive tests adopting different model assumptions, and obtaining radial velocities with different methods including RV measurements from an analysis of CCF bisector and measuring RVs from LSD profiles produced by an independent pipeline \citep[submitted]{donati2020}. All RM model fits persistently prefer a $\lambda$ value consistent 
with aligned rotation and orbital angular momenta.

\section{Conclusions}
\label{sec:conclusions}

   \begin{enumerate}
   
        \item We present observations of a transit of the recently detected planet of the nearby young M1 star \aumic\ with a resolved edge-on debris disk with the SPIRou high resolution near infrared spectropolarimeter at CFHT and the iSHELL high resolution near infrared spectrograph at IRTF .

      \item We cross-correlate the spectra with numerical masks and employ the CCF matching method to obtain radial velocities of \aumic\ with $\sim$5-\mps\ precision. 
      
      \item We obtain Stokes I and V spectra of \aumic\ and perform a LSD analysis to obtain average Stokes I and V profiles, and strongly detect a Zeeman signature in the Stokes V profile.  The corresponding mean longitudinal magnetic field is $\overline{B}_\ell=46.3\pm0.7$~G and varies at a global rate of $d{B_\ell}/dt=-108.7\pm7.7$~G/d. 
      
      \item We use the correlated variability of the longitudinal magnetic field and radial velocity, with a scaling factor $a=-1.34\pm0.12$~\mpspG\ to empirically correct a linear RV trend of $145\pm{17}$~\mpspd. This trend is consistent with the slope of $149\pm{9}$~\mpspd\ found in our RM analysis and compatible with the expected levels of RV jitter of \aumic\ in the nIR \citep{Bailey2012}.
      
      \item We fit a classical Rossiter-McLaughlin effect model to the SPIRou and iSHELL data, and find a sky projected spin-orbit obliquity angle for \aumicb\ of $\lambda=0^{+18}_{-15}$ degrees.
      
      \item \aumicb\ is therefore on a prograde and closely aligned orbit, which is evidence that the planet likely formed in the protoplanetary disk that evolved to the current \aumic\ debris disk, provided that the star-disk-planet components of the system share the same angular momentum orientation. 

   \end{enumerate}

\begin{acknowledgements}
      The authors wish to recognize and acknowledge the very significant cultural role and reverence that the summit of Maunakea has always had within the indigenous Hawaiian community. We are most fortunate to have the opportunity to conduct observations from this mountain. 
      We acknowledge funding from the French National Research Agency (ANR) under contract number ANR-18-CE31-0019 (SPlaSH) and also in the framework of the Investissements dAvenir program (ANR-15-IDEX-02), through the funding of the ”Origin of Life” project of the Univ. Grenoble-Alpes. We also acknowledge funding from the European Research Council (ERC) under the H2020 research \& innovation programme (grant agreement 740651 NewWorlds).  SPIRou project is funded by  the IDEX initiative at UFTMP, UPS, the DIM-ACAV programme in Region Ile de France, the MIDEX initiative atAMU, the Labex@OSUG2020 programme, UGA, INSU/CNRS, CFI, CFHT, LNA, CAUP and DIAS. We are also grateful for generous amounts of in-kind manpower allocated to SPIRou by OMP/IRAP, OHP/LAM, IPAG, CFHT, NRC-H, UdeM, UL, OG, LNA and ASIAA. P.P. acknowledges support from the NASA Exoplanet Exploration Program and the National Science Foundation (Astronomy and Astrophysics grant 1716202). J.M. acknowledges support from Eric Stempels and  Nikolai Piskunov of the VALD team at Uppsala University. J.M. also thanks Benjamin Tessore (IPAG), Eric Josselin (LUPM) and Agnès Lèbre (LUPM) for their contribution to the Montpellier VALD local mirror and Bertrand Plez (LUPM) for his assistance with MARCS model atmospheres.

\end{acknowledgements}

%
%

\bibliographystyle{aa}

\bibliography{aumicb}

\begin{thebibliography}{34}
\expandafter\ifx\csname natexlab\endcsname\relax\def\natexlab#1{#1}\fi

\bibitem[{{Afram} \& {Berdyugina}(2019)}]{Afram2019}
{Afram}, N. \& {Berdyugina}, S.~V. 2019, \aap, 629, A83

\bibitem[{{Artigau} {et~al.}(2014){Artigau}, {Astudillo-Defru}, {Delfosse},
  {Bouchy}, {Bonfils}, {Lovis}, {Pepe}, {Moutou}, {Donati}, {Doyon}, \&
  {Malo}}]{artigau14}
{Artigau}, {\'E}., {Astudillo-Defru}, N., {Delfosse}, X., {et~al.} 2014, in
  Society of Photo-Optical Instrumentation Engineers (SPIE) Conference Series,
  Vol. 9149, \procspie, 914905

\bibitem[{{Bagnulo} {et~al.}(2009){Bagnulo}, {Landolfi}, {Landstreet}, {Land i
  Degl'Innocenti}, {Fossati}, \& {Sterzik}}]{bagnulo2009}
{Bagnulo}, S., {Landolfi}, M., {Landstreet}, J.~D., {et~al.} 2009, \pasp, 121,
  993

\bibitem[{{Bailey} {et~al.}(2012){Bailey}, {White}, {Blake}, {Charbonneau},
  {Barman}, {Tanner}, \& {Torres}}]{Bailey2012}
{Bailey}, John~I., I., {White}, R.~J., {Blake}, C.~H., {et~al.} 2012, \apj,
  749, 16

\bibitem[{{Berdyugina} {et~al.}(2008){Berdyugina}, {Berdyugin}, {Fluri}, \&
  {Piirola}}]{Berdyugina2008}
{Berdyugina}, S.~V., {Berdyugin}, A.~V., {Fluri}, D.~M., \& {Piirola}, V. 2008,
  \apjl, 673, L83

\bibitem[{{Cale} {et~al.}(2019){Cale}, {Plavchan}, {LeBrun}, {Gagn{\'e}},
  {Gao}, {Tanner}, {Beichman}, {Xuesong Wang}, {Gaidos}, {Teske}, {Ciardi},
  {Vasisht}, {Kane}, \& {von Braun}}]{Cale2019}
{Cale}, B., {Plavchan}, P., {LeBrun}, D., {et~al.} 2019, \aj, 158, 170

\bibitem[{{Claret} \& {Bloemen}(2011)}]{claret2011}
{Claret}, A. \& {Bloemen}, S. 2011, VizieR Online Data Catalog, J/A+A/529/A75

\bibitem[{Cuillandre {et~al.}(2004)Cuillandre, Magnier, Isani, Sabin, Knight,
  Kras, \& Lai}]{Cuillandre2004}
Cuillandre, J.-C., Magnier, E.~A., Isani, S., {et~al.} 2004, in Scientific
  Detectors for Astronomy, ed. P.~Amico, J.~W. Beletic, \& J.~E. Beletic
  (Dordrecht: Springer Netherlands), 287--298

\bibitem[{{David} {et~al.}(2016){David}, {Hillenbrand}, {Petigura},
  {Carpenter}, {Crossfield}, {Hinkley}, {Ciardi}, {Howard}, {Isaacson}, {Cody},
  {Schlieder}, {Beichman}, \& {Barenfeld}}]{David2016}
{David}, T.~J., {Hillenbrand}, L.~A., {Petigura}, E.~A., {et~al.} 2016, \nat,
  534, 658

\bibitem[{{Dawson} \& {Johnson}(2018)}]{DawsonAndJohnson2018}
{Dawson}, R.~I. \& {Johnson}, J.~A. 2018, \araa, 56, 175

\bibitem[{{Donati} {et~al.}(2020){Donati}, {Kouach}, {Moutou}, {Doyon},
  {Delfosse}, {Artigau}, {Baratchart}, {Lacombe}, \& {Barrick}}]{donati2020}
{Donati}, J.~F., {Kouach}, D., {Moutou}, C., {et~al.} 2020, Submitted to \mnras

\bibitem[{{Donati} {et~al.}(2016){Donati}, {Moutou}, {Malo}, {Baruteau}, {Yu},
  {H{\'e}brard}, {Hussain}, {Alencar}, {M{\'e}nard}, {Bouvier}, {Petit},
  {Takami}, {Doyon}, \& {Cameron}}]{donati2016}
{Donati}, J.~F., {Moutou}, C., {Malo}, L., {et~al.} 2016, \nat, 534, 662

\bibitem[{{Donati} {et~al.}(1997){Donati}, {Semel}, {Carter}, {Rees}, \&
  {Collier Cameron}}]{donati97}
{Donati}, J.~F., {Semel}, M., {Carter}, B.~D., {Rees}, D.~E., \& {Collier
  Cameron}, A. 1997, \mnras, 291, 658

\bibitem[{{Foreman-Mackey} {et~al.}(2013){Foreman-Mackey}, {Hogg}, {Lang}, \&
  {Goodman}}]{foreman2013}
{Foreman-Mackey}, D., {Hogg}, D.~W., {Lang}, D., \& {Goodman}, J. 2013, \pasp,
  125, 306

\bibitem[{{Gaia Collaboration}(2018)}]{gaiadr22018}
{Gaia Collaboration}. 2018, VizieR Online Data Catalog, I/345

\bibitem[{{Gaidos} {et~al.}(2014){Gaidos}, {Mann}, {L{\'e}pine}, {Buccino},
  {James}, {Ansdell}, {Petrucci}, {Mauas}, \& {Hilton}}]{gaidos2014}
{Gaidos}, E., {Mann}, A.~W., {L{\'e}pine}, S., {et~al.} 2014, \mnras, 443, 2561

\bibitem[{{Gustafsson} {et~al.}(2008){Gustafsson}, {Edvardsson}, {Eriksson},
  {J{\o}rgensen}, {Nordlund}, \& {Plez}}]{Gustafsson2008}
{Gustafsson}, B., {Edvardsson}, B., {Eriksson}, K., {et~al.} 2008, \aap, 486,
  951

\bibitem[{{Horne}(1986)}]{horne86}
{Horne}, K. 1986, \pasp, 98, 609

\bibitem[{{Kalas} {et~al.}(2004){Kalas}, {Liu}, \& {Matthews}}]{kalas2004}
{Kalas}, P., {Liu}, M.~C., \& {Matthews}, B.~C. 2004, Science, 303, 1990

\bibitem[{{Lagrange} {et~al.}(2009){Lagrange}, {Kasper}, {Boccaletti},
  {Chauvin}, {Gratadour}, {Fusco}, {Ehrenreich}, {Apai}, {Mouillet}, \&
  {Rouan}}]{Lagrange2009}
{Lagrange}, A.~M., {Kasper}, M., {Boccaletti}, A., {et~al.} 2009, \aap, 506,
  927

\bibitem[{{Lavail} {et~al.}(2018){Lavail}, {Kochukhov}, \& {Wade}}]{Lavail2018}
{Lavail}, A., {Kochukhov}, O., \& {Wade}, G.~A. 2018, \mnras, 479, 4836

\bibitem[{{Mamajek} \& {Bell}(2014)}]{mamjek2014}
{Mamajek}, E.~E. \& {Bell}, C. P.~M. 2014, \mnras, 445, 2169

\bibitem[{{Mann} {et~al.}(2016){Mann}, {Newton}, {Rizzuto}, {Irwin}, {Feiden},
  {Gaidos}, {Mace}, {Kraus}, {James}, {Ansdell}, {Charbonneau}, {Covey},
  {Ireland}, {Jaffe}, {Johnson}, {Kidder}, \& {Vanderburg}}]{Mann2016}
{Mann}, A.~W., {Newton}, E.~R., {Rizzuto}, A.~C., {et~al.} 2016, \aj, 152, 61

\bibitem[{{Morin} {et~al.}(2010){Morin}, {Donati}, {Petit}, {Delfosse},
  {Forveille}, \& {Jardine}}]{Morin2010}
{Morin}, J., {Donati}, J.~F., {Petit}, P., {et~al.} 2010, \mnras, 407, 2269

\bibitem[{{Moutou} {et~al.}(2020){Moutou}, {Dalal}, {Donati}, {Martioli},
  {Folsom}, \& et~al}]{moutou2020}
{Moutou}, C., {Dalal}, S., {Donati}, J.-F., {et~al.} 2020, Accepted for
  publication at \aap

\bibitem[{{Newton} {et~al.}(2019){Newton}, {Mann}, {Tofflemire}, {Pearce},
  {Rizzuto}, {Vanderburg}, {Martinez}, {Wang}, {Ruffio}, {Kraus}, {Johnson},
  {Thao}, {Wood}, {Rampalli}, {Nielsen}, {Collins}, {Dragomir}, {Hellier},
  {Anderson}, {Barclay}, {Brown}, {Feiden}, {Hart}, {Isopi}, {Kielkopf},
  {Mallia}, {Nelson}, {Rodriguez}, {Stockdale}, {Waite}, {Wright}, {Lissauer},
  {Ricker}, {Vanderspek}, {Latham}, {Seager}, {Winn}, {Jenkins}, {Bouma},
  {Burke}, {Davies}, {Fausnaugh}, {Li}, {Morris}, {Mukai}, {Villase{\~n}or},
  {Villeneuva}, {De Rosa}, {Macintosh}, {Mengel}, {Okumura}, \&
  {Wittenmyer}}]{Newton2019}
{Newton}, E.~R., {Mann}, A.~W., {Tofflemire}, B.~M., {et~al.} 2019, \apjl, 880,
  L17

\bibitem[{{Ohta} {et~al.}(2005){Ohta}, {Taruya}, \& {Suto}}]{ohta2005}
{Ohta}, Y., {Taruya}, A., \& {Suto}, Y. 2005, \apj, 622, 1118

\bibitem[{{Pepe} {et~al.}(2002){Pepe}, {Mayor}, {Galland}, {Naef}, {Queloz},
  {Santos}, {Udry}, \& {Burnet}}]{pepe2002}
{Pepe}, F., {Mayor}, M., {Galland}, F., {et~al.} 2002, \aap, 388, 632

\bibitem[{{Piskunov} {et~al.}(1995){Piskunov}, {Kupka}, {Ryabchikova}, {Weiss},
  \& {Jeffery}}]{piskunov1995}
{Piskunov}, N.~E., {Kupka}, F., {Ryabchikova}, T.~A., {Weiss}, W.~W., \&
  {Jeffery}, C.~S. 1995, \aaps, 112, 525

\bibitem[{{Plavchan} {et~al.}(2020){Plavchan}, {Barclay}, {Gagn{\'e}}, {Gao},
  {Cale}, {Matzko}, {Dragomir}, {Quinn}, {Feliz}, {Stassun}, {Crossfield},
  {Berardo}, {Latham}, {Tieu}, {Anglada-Escud{\'e}}, {Ricker}, {Vanderspek},
  {Seager}, {Winn}, {Jenkins}, {Rinehart}, {Krishnamurthy}, {Dynes}, {Doty},
  {Adams}, {Afanasev}, {Beichman}, {Bottom}, {Bowler}, {Brinkworth}, {Brown},
  {Cancino}, {Ciardi}, {Clampin}, {Clark}, {Collins}, {Davison},
  {Foreman-Mackey}, {Furlan}, {Gaidos}, {Geneser}, {Giddens}, {Gilbert},
  {Hall}, {Hellier}, {Henry}, {Horner}, {Howard}, {Huang}, {Huber}, {Kane},
  {Kenworthy}, {Kielkopf}, {Kipping}, {Klenke}, {Kruse}, {Latouf}, {Lowrance},
  {Mennesson}, {Mengel}, {Mills}, {Morton}, {Narita}, {Newton}, {Nishimoto},
  {Okumura}, {Palle}, {Pepper}, {Quintana}, {Roberge}, {Roccatagliata},
  {Schlieder}, {Tanner}, {Teske}, {Tinney}, {Vanderburg}, {von Braun}, {Walp},
  {Wang}, {Wang}, {Weigand }, {White}, {Wittenmyer}, {Wright}, {Youngblood},
  {Zhang}, \& {Zilberman}}]{plavchan2020Natur}
{Plavchan}, P., {Barclay}, T., {Gagn{\'e}}, J., {et~al.} 2020, \nat, 582, 497

\bibitem[{{Rayner} {et~al.}(2016){Rayner}, {Tokunaga}, {Jaffe}, {Bonnet},
  {Ching}, {Connelley}, {Kokubun}, {Lockhart}, \& {Warmbier}}]{Rayner2016}
{Rayner}, J., {Tokunaga}, A., {Jaffe}, D., {et~al.} 2016, in Society of
  Photo-Optical Instrumentation Engineers (SPIE) Conference Series, Vol. 9908,
  \procspie, 990884

\bibitem[{{Torres} {et~al.}(2006){Torres}, {Quast}, {da Silva}, {de La Reza},
  {Melo}, \& {Sterzik}}]{torres2006}
{Torres}, C.~A.~O., {Quast}, G.~R., {da Silva}, L., {et~al.} 2006, \aap, 460,
  695

\bibitem[{{Triaud}(2018)}]{Triaud2018}
{Triaud}, A. H.~M.~J. 2018, {The Rossiter-McLaughlin Effect in Exoplanet
  Research}, 2

\bibitem[{{White} {et~al.}(2015){White}, {Schaefer}, {Ten Brummelaar},
  {Farrington}, {McAlister}, {Ridgway}, {sturmann}, {Sturmann}, \&
  {Turner}}]{russel2015}
{White}, R.~J., {Schaefer}, G., {Ten Brummelaar}, T., {et~al.} 2015, in
  American Astronomical Society Meeting Abstracts, Vol. 225, American
  Astronomical Society Meeting Abstracts \#225, 348.12

\end{thebibliography}

\begin{appendix}

\section{Radial velocity data}
\label{app:rvdata}

\begin{longtable}{cc}
\caption{Radial velocity data of \aumic\ measured by SPIRou subtracted the mean RV of -4.3917~\kps.}\\
\label{tab:aumicspirourvdata}
BJD & RV \\
    & \mps\ \\
\hline
 2458651.9294631 & $13.4\pm13.3$ \\
 2458651.9313410 & $15.0\pm12.7$ \\
 2458651.9331425 & $22.9\pm9.6$ \\
 2458651.9349485 & $12.6\pm10.5$ \\
 2458651.9367580 & $11.1\pm7.6$ \\
 2458651.9386249 & $2.7\pm7.7$ \\
 2458651.9404333 & $4.2\pm7.3$ \\
 2458651.9422373 & $3.6\pm8.5$ \\
 2458651.9440437 & $8.4\pm8.1$ \\
 2458651.9459171 & $9.3\pm6.3$ \\
 2458651.9477209 & $11.1\pm6.6$ \\
 2458651.9495294 & $10.4\pm8.3$ \\
 2458651.9513325 & $-2.3\pm7.6$ \\
 2458651.9532062 & $-5.0\pm5.4$ \\
 2458651.9550775 & $-7.0\pm5.2$ \\
 2458651.9568821 & $-3.1\pm4.8$ \\
 2458651.9586890 & $-4.2\pm4.9$ \\
 2458651.9605578 & $2.2\pm2.3$ \\
 2458651.9624290 & $0.3\pm3.8$ \\
 2458651.9642418 & $-3.1\pm3.4$ \\
 2458651.9660419 & $-2.6\pm3.9$ \\
 2458651.9678465 & $-7.0\pm4.8$ \\
 2458651.9697194 & $-1.8\pm4.6$ \\
 2458651.9715249 & $-3.3\pm2.6$ \\
 2458651.9733311 & $-6.9\pm3.6$ \\
 2458651.9751368 & $-14.3\pm4.2$ \\
 2458651.9770093 & $-10.4\pm4.5$ \\
 2458651.9788159 & $-6.9\pm4.2$ \\
 2458651.9806213 & $3.1\pm5.1$ \\
 2458651.9824924 & $-3.2\pm2.8$ \\
 2458651.9843032 & $-6.7\pm4.9$ \\
 2458651.9861111 & $-12.5\pm4.6$ \\
 2458651.9879170 & $-11.9\pm4.8$ \\
 2458651.9897810 & $-11.2\pm3.9$ \\
 2458651.9916527 & $-10.7\pm3.3$ \\
 2458651.9934624 & $-2.9\pm1.9$ \\
 2458651.9952670 & $-0.7\pm2.6$ \\
 2458651.9970739 & $-5.8\pm2.7$ \\
 2458651.9989478 & $-9.0\pm3.0$ \\
 2458652.0007502 & $-11.5\pm3.2$ \\
 2458652.0025548 & $-18.7\pm3.8$ \\
 2458652.0044254 & $-2.1\pm3.6$ \\
 2458652.0062317 & $-12.9\pm4.0$ \\
 2458652.0080383 & $-9.7\pm4.3$ \\
 2458652.0098456 & $-22.6\pm2.6$ \\
 2458652.0116529 & $-10.5\pm3.3$ \\
 2458652.0135269 & $-16.5\pm3.9$ \\
 2458652.0153300 & $-14.2\pm3.8$ \\
 2458652.0171348 & $-17.7\pm2.8$ \\
 2458652.0189428 & $-16.0\pm6.0$ \\
 2458652.0208140 & $-23.7\pm6.1$ \\
 2458652.0226186 & $-23.0\pm4.7$ \\
 2458652.0244869 & $-9.2\pm4.3$ \\
 2458652.0263629 & $-6.7\pm2.9$ \\
 2458652.0282321 & $-10.0\pm4.2$ \\
 2458652.0300386 & $-15.0\pm8.3$ \\
 2458652.0318420 & $-7.0\pm9.7$ \\
 2458652.0336501 & $-0.2\pm6.8$ \\
 2458652.0355208 & $6.4\pm5.3$ \\
 2458652.0373361 & $-0.6\pm6.5$ \\
 2458652.0391326 & $-10.8\pm7.9$ \\
 2458652.0409384 & $-15.0\pm7.9$ \\
 2458652.0428097 & $-21.7\pm4.8$ \\
 2458652.0446176 & $-27.2\pm4.5$ \\
 2458652.0464214 & $-25.6\pm4.9$ \\
 2458652.0482933 & $-17.3\pm5.0$ \\
 2458652.0501654 & $-25.0\pm2.5$ \\
 2458652.0519706 & $-27.7\pm2.6$ \\
 2458652.0537768 & $-28.9\pm2.8$ \\
 2458652.0555851 & $-28.1\pm3.4$ \\
 2458652.0574541 & $-19.2\pm4.3$ \\
 2458652.0592612 & $-14.4\pm4.7$ \\
 2458652.0610671 & $-9.4\pm4.9$ \\
 2458652.0629438 & $-11.6\pm5.1$ \\
 2458652.0648082 & $-3.9\pm5.9$ \\
 2458652.0666136 & $0.1\pm3.3$ \\
 2458652.0684188 & $4.2\pm3.8$ \\
 2458652.0702908 & $0.7\pm4.0$ \\
 2458652.0721621 & $4.6\pm3.9$ \\
 2458652.0739677 & $6.7\pm5.1$ \\
 2458652.0757738 & $2.7\pm5.8$ \\
 2458652.0776455 & $-3.5\pm4.3$ \\
 2458652.0795174 & $7.8\pm4.4$ \\
 2458652.0813268 & $20.3\pm3.5$ \\
 2458652.0831301 & $26.9\pm3.9$ \\
 2458652.0849380 & $28.5\pm2.2$ \\
 2458652.0868071 & $24.5\pm3.4$ \\
 2458652.0886184 & $20.8\pm3.7$ \\
 2458652.0904247 & $28.0\pm5.0$ \\
 2458652.0922889 & $18.8\pm5.4$ \\
 2458652.0941018 & $25.1\pm5.7$ \\
 2458652.0959712 & $23.5\pm5.5$ \\
 2458652.0977723 & $25.1\pm5.5$ \\
 2458652.0996440 & $22.1\pm5.0$ \\
 2458652.1015193 & $19.5\pm4.6$ \\
 2458652.1033195 & $27.8\pm3.0$ \\
 2458652.1051266 & $25.2\pm2.8$ \\
 2458652.1069325 & $24.8\pm3.7$ \\
 2458652.1088048 & $22.0\pm3.2$ \\
 2458652.1106092 & $16.2\pm3.2$ \\
 2458652.1124186 & $12.2\pm3.2$ \\
 2458652.1142236 & $11.6\pm3.3$ \\
 2458652.1160988 & $-0.9\pm4.1$ \\
 2458652.1179010 & $1.1\pm4.1$ \\
 2458652.1197077 & $3.4\pm5.4$ \\
 2458652.1215767 & $13.4\pm5.2$ \\
 2458652.1233840 & $12.3\pm4.9$ \\
 2458652.1251891 & $12.8\pm4.8$ \\
 2458652.1269964 & $9.0\pm6.0$ \\
 2458652.1288033 & $5.6\pm7.0$ \\
 2458652.1306100 & $1.3\pm7.8$ \\
 2458652.1324193 & $9.0\pm6.9$ \\
 2458652.1342873 & $6.2\pm4.7$ \\
 2458652.1361572 & $11.4\pm3.3$ \\
 2458652.1379697 & $7.7\pm4.4$ \\
 2458652.1397685 & $10.8\pm6.4$ \\
\hline
\end{longtable}

\begin{longtable}{cc}
\caption{Radial velocity data of \aumic\ measured by iSHELL.}\\
\label{tab:aumicishellrvdata}
BJD & RV \\
    & \mps\ \\
\hline
 2458651.9699598 & $19.0\pm20.5$ \\
 2458651.9715484 & $2.1\pm21.9$ \\
 2458651.9731365 & $17.4\pm27.8$ \\
 2458651.9747251 & $31.6\pm24.6$ \\
 2458651.9763132 & $13.1\pm22.6$ \\
 2458651.9779024 & $15.8\pm23.9$ \\
 2458651.9794905 & $23.2\pm22.8$ \\
 2458651.9810791 & $29.0\pm23.2$ \\
 2458651.9826683 & $19.1\pm23.1$ \\
 2458651.9842570 & $5.0\pm25.3$ \\
 2458651.9858462 & $41.2\pm21.2$ \\
 2458651.9874342 & $16.8\pm19.7$ \\
 2458651.9890229 & $13.9\pm20.0$ \\
 2458651.9906121 & $18.9\pm21.3$ \\
 2458651.9922002 & $21.3\pm18.4$ \\
 2458651.9937888 & $16.9\pm17.6$ \\
 2458651.9953769 & $23.9\pm24.3$ \\
 2458651.9969661 & $16.5\pm23.0$ \\
 2458651.9985547 & $23.0\pm24.5$ \\
 2458652.0001434 & $17.3\pm17.6$ \\
 2458652.0017320 & $10.8\pm23.2$ \\
 2458652.0033206 & $6.0\pm20.8$ \\
 2458652.0049087 & $-1.5\pm23.5$ \\
 2458652.0064973 & $4.7\pm20.0$ \\
 2458652.0080860 & $-10.2\pm20.1$ \\
 2458652.0096740 & $11.0\pm19.3$ \\
 2458652.0112633 & $19.2\pm21.6$ \\
 2458652.0128519 & $-6.3\pm15.5$ \\
 2458652.0144400 & $8.1\pm20.7$ \\
 2458652.0160280 & $2.6\pm26.8$ \\
 2458652.0176172 & $1.2\pm21.6$ \\
 2458652.0192059 & $-28.9\pm23.7$ \\
 2458652.0207945 & $1.1\pm23.7$ \\
 2458652.0223837 & $-6.4\pm21.1$ \\
 2458652.0239724 & $27.6\pm24.9$ \\
 2458652.0255616 & $-2.9\pm16.3$ \\
 2458652.0271496 & $32.4\pm20.4$ \\
 2458652.0287383 & $-2.8\pm16.5$ \\
 2458652.0303263 & $-22.7\pm16.1$ \\
 2458652.0319150 & $-10.7\pm25.1$ \\
 2458652.0335030 & $-11.8\pm20.4$ \\
 2458652.0350917 & $-3.6\pm27.3$ \\
 2458652.0366797 & $-10.2\pm17.8$ \\
 2458652.0382678 & $-8.7\pm18.5$ \\
 2458652.0398559 & $-12.5\pm22.6$ \\
 2458652.0414439 & $-6.9\pm17.2$ \\
 2458652.0430326 & $-13.1\pm20.2$ \\
\hline
\end{longtable}

\section{Longitudinal magnetic field data}
\label{app:Bldata}

\begin{longtable}{cc}
\caption{Longitudinal magnetic field data of \aumic\ measured by SPIRou. }\\
\label{tab:aumicspiroublongdata}
BJD & $B_\ell$ \\
    & G \\
\hline
2458651.9315144 & $72.2\pm10.3$ \\
 2458651.9333381 & $69.4\pm10.5$ \\
 2458651.9351591 & $67.1\pm10.2$ \\
 2458651.9369818 & $61.0\pm10.0$ \\
 2458651.9388040 & $57.6\pm9.4$ \\
 2458651.9406254 & $54.2\pm8.7$ \\
 2458651.9424485 & $52.8\pm8.3$ \\
 2458651.9442704 & $55.3\pm8.2$ \\
 2458651.9460934 & $61.0\pm8.2$ \\
 2458651.9479156 & $61.7\pm8.1$ \\
 2458651.9497379 & $61.8\pm7.9$ \\
 2458651.9515770 & $60.6\pm7.8$ \\
 2458651.9534152 & $59.4\pm7.8$ \\
 2458651.9552543 & $57.0\pm7.8$ \\
 2458651.9570922 & $56.2\pm8.0$ \\
 2458651.9589301 & $55.3\pm8.0$ \\
 2458651.9607700 & $52.8\pm8.0$ \\
 2458651.9626082 & $52.9\pm7.9$ \\
 2458651.9644304 & $55.0\pm7.8$ \\
 2458651.9662530 & $56.6\pm7.9$ \\
 2458651.9680738 & $59.8\pm7.8$ \\
 2458651.9698961 & $56.6\pm7.7$ \\
 2458651.9717187 & $56.1\pm7.8$ \\
 2458651.9735412 & $55.1\pm7.9$ \\
 2458651.9753639 & $54.1\pm8.0$ \\
 2458651.9771864 & $51.9\pm8.1$ \\
 2458651.9790253 & $51.8\pm8.0$ \\
 2458651.9808488 & $53.3\pm7.9$ \\
 2458651.9826726 & $53.7\pm7.8$ \\
 2458651.9844965 & $53.4\pm7.7$ \\
 2458651.9863187 & $53.5\pm7.7$ \\
 2458651.9881561 & $50.7\pm7.6$ \\
 2458651.9899939 & $50.6\pm7.5$ \\
 2458651.9918314 & $50.1\pm7.4$ \\
 2458651.9936546 & $49.2\pm7.3$ \\
 2458651.9954784 & $48.8\pm7.3$ \\
 2458651.9973003 & $46.3\pm7.5$ \\
 2458651.9991223 & $48.2\pm7.6$ \\
 2458652.0009602 & $47.7\pm7.8$ \\
 2458652.0027812 & $47.1\pm7.8$ \\
 2458652.0046032 & $45.9\pm7.7$ \\
 2458652.0064259 & $44.6\pm7.7$ \\
 2458652.0082327 & $48.8\pm7.6$ \\
 2458652.0100565 & $50.2\pm7.4$ \\
 2458652.0118794 & $51.3\pm7.4$ \\
 2458652.0137018 & $52.8\pm7.3$ \\
 2458652.0155242 & $54.1\pm7.3$ \\
 2458652.0173460 & $55.7\pm7.4$ \\
 2458652.0191682 & $54.3\pm7.4$ \\
 2458652.0210062 & $53.6\pm7.5$ \\
 2458652.0228612 & $53.6\pm7.5$ \\
 2458652.0247157 & $51.1\pm7.7$ \\
 2458652.0265707 & $51.4\pm7.8$ \\
 2458652.0284095 & $52.3\pm7.8$ \\
 2458652.0302313 & $48.5\pm7.8$ \\
 2458652.0320535 & $44.4\pm7.7$ \\
 2458652.0338779 & $43.3\pm7.7$ \\
 2458652.0357005 & $42.1\pm7.5$ \\
 2458652.0375226 & $40.8\pm7.5$ \\
 2458652.0393448 & $40.8\pm7.4$ \\
 2458652.0411652 & $39.7\pm7.5$ \\
 2458652.0429874 & $39.2\pm7.5$ \\
 2458652.0448261 & $40.6\pm7.6$ \\
 2458652.0466651 & $40.6\pm7.5$ \\
 2458652.0485033 & $41.0\pm7.5$ \\
 2458652.0503422 & $41.9\pm7.6$ \\
 2458652.0521651 & $45.8\pm7.6$ \\
 2458652.0539873 & $48.2\pm7.8$ \\
 2458652.0558099 & $49.6\pm7.9$ \\
 2458652.0576325 & $49.4\pm7.9$ \\
 2458652.0594722 & $49.0\pm8.0$ \\
 2458652.0613107 & $47.1\pm8.1$ \\
 2458652.0631488 & $43.6\pm8.1$ \\
 2458652.0649867 & $40.4\pm7.9$ \\
 2458652.0668235 & $40.7\pm7.8$ \\
 2458652.0686619 & $38.4\pm7.8$ \\
 2458652.0705005 & $38.2\pm8.0$ \\
 2458652.0723392 & $42.3\pm8.4$ \\
 2458652.0741779 & $39.9\pm8.6$ \\
 2458652.0760167 & $44.0\pm8.7$ \\
 2458652.0778565 & $43.2\pm8.6$ \\
 2458652.0796956 & $41.1\pm8.5$ \\
 2458652.0815187 & $38.2\pm8.5$ \\
 2458652.0833411 & $37.4\pm8.4$ \\
 2458652.0851640 & $38.2\pm8.5$ \\
 2458652.0869876 & $37.8\pm8.5$ \\
 2458652.0888254 & $38.1\pm8.5$ \\
 2458652.0906491 & $40.1\pm8.4$ \\
 2458652.0924873 & $40.9\pm8.2$ \\
 2458652.0943242 & $40.1\pm8.1$ \\
 2458652.0961630 & $37.9\pm7.9$ \\
 2458652.0980173 & $37.2\pm7.8$ \\
 2458652.0998544 & $38.1\pm7.9$ \\
 2458652.1016930 & $40.0\pm8.0$ \\
 2458652.1035151 & $39.6\pm8.0$ \\
 2458652.1053364 & $40.5\pm8.0$ \\
 2458652.1071589 & $42.8\pm7.9$ \\
 2458652.1089819 & $44.6\pm7.8$ \\
 2458652.1108047 & $48.9\pm7.8$ \\
 2458652.1126282 & $50.0\pm7.8$ \\
 2458652.1144511 & $52.8\pm7.7$ \\
 2458652.1162734 & $51.9\pm7.6$ \\
 2458652.1181116 & $51.0\pm7.6$ \\
 2458652.1199329 & $49.0\pm7.5$ \\
 2458652.1217550 & $47.2\pm7.5$ \\
 2458652.1235772 & $44.2\pm7.5$ \\
 2458652.1253838 & $41.4\pm7.5$ \\
 2458652.1271903 & $39.2\pm7.5$ \\
 2458652.1289979 & $34.9\pm7.5$ \\
 2458652.1308206 & $34.2\pm7.5$ \\
 2458652.1326591 & $32.7\pm7.6$ \\
 2458652.1344990 & $33.3\pm7.6$ \\
 2458652.1363363 & $30.6\pm7.7$ \\
\hline
\end{longtable}

\section{Posterior distributions}
\label{app:posteriordistributions}

  \begin{figure*}
   \centering
   \includegraphics[width=0.9\hsize]{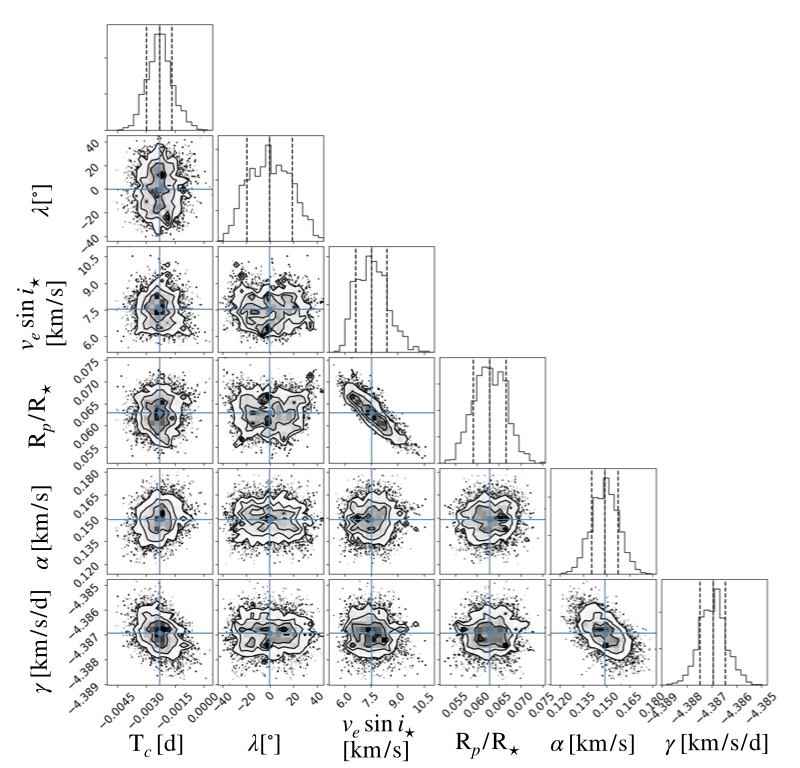}
      \caption{Pairs plot showing the MCMC samples and posterior distributions for the six free parameters presented in Table \ref{tab:aumicfitparams}. The contours mark the 1$\sigma$, 2$\sigma$, and 3$\sigma$ regions of the distribution. The gray scale shades illustrate the density of samples, where darker means denser. The blue crosses indicate the best fit values for each parameter and the dashed vertical lines in the projected distributions indicate the median value and the 1-$\sigma$ uncertainty (34\% on each side of the median).
         }
        \label{fig:rmfit-to-spirourv_pairsplot}
  \end{figure*}

 \begin{figure*}
   \centering
   \includegraphics[width=0.9\hsize]{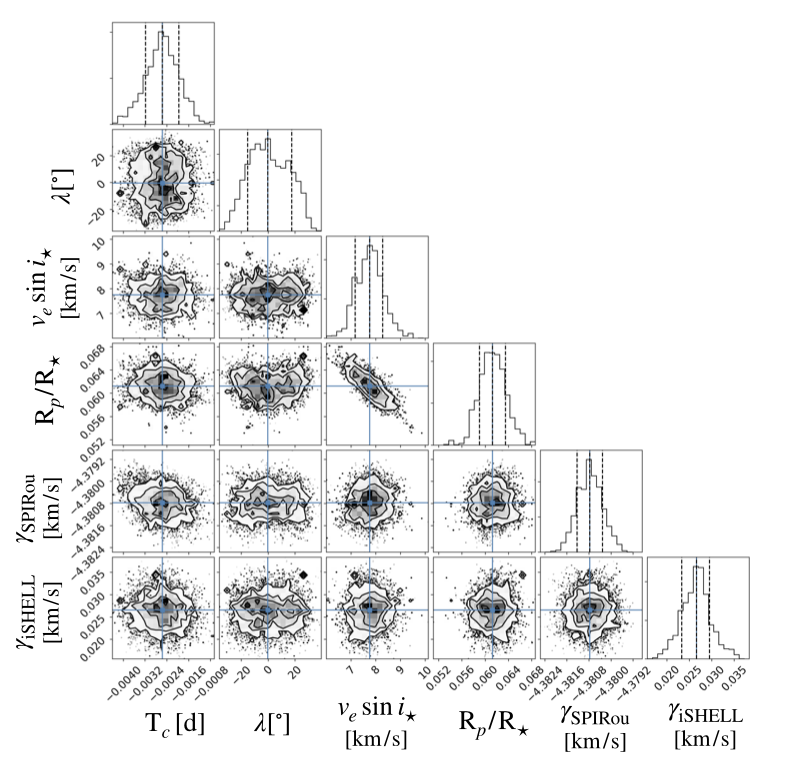}
      \caption{Pairs plot showing the MCMC samples and posterior distributions for the six free parameters presented in Table \ref{tab:aumicfinalfitparams}. The contours mark the 1$\sigma$, 2$\sigma$, and 3$\sigma$ regions of the distribution. The gray scale shades illustrate the density of samples, where darker means denser. The blue crosses indicate the best fit values for each parameter and the dashed vertical lines in the projected distributions indicate the median value and the 1$\sigma$ uncertainty (34\% on each side of the median).
         }
        \label{fig:rmfit-to-spirou+ishell_pairsplot}
 \end{figure*}  

\end{appendix}

\end{document}